\documentclass[journal=jacsat,manuscript=article]{achemso} %,layout=twocolumn
\usepackage[version=3]{mhchem} % Formula subscripts using \ce{}
\usepackage{amsmath,amssymb}
\usepackage{hyperref}
\hypersetup{
    colorlinks=true,
    linkcolor=blue,
    citecolor=red,
    filecolor=magenta,      
    urlcolor=cyan}
\usepackage[font=small]{caption} %footnotesize
\SectionNumbersOn %to number the different sections

\makeatletter
\let\l@addto@macro\relax
\makeatother
\usepackage[fontsize=11pt]{scrextend}

\author{Yuchun Zhu}
\affiliation{Institute of Physics, Swiss Federal Institute of Technology (EPFL), CH-1015 Lausanne, Switzerland}
\email{yuchun.zhu@epfl.ch}
\author{Elena Losero}
\affiliation{Institute of Physics, Swiss Federal Institute of Technology (EPFL), CH-1015 Lausanne, Switzerland}
\alsoaffiliation{Division of Quantum Metrology and Nanotechnologies, Istituto Nazionale di Ricerca Metrologica (INRiM), 10135 Torino, Italy}
\author{Christophe Galland}
\affiliation{Institute of Physics, Swiss Federal Institute of Technology (EPFL), CH-1015 Lausanne, Switzerland}
\alsoaffiliation{Center for Quantum Science and Engineering, EPFL, Lausanne, Switzerland}
\author{Valentin Goblot}
\affiliation{Institute of Physics, Swiss Federal Institute of Technology (EPFL), CH-1015 Lausanne, Switzerland}
\alsoaffiliation{Center for Quantum Science and Engineering, EPFL, Lausanne, Switzerland}

\title{Simulation of ODMR Spectra from Nitrogen-Vacancy Ensembles in Diamond for Electric Field Sensing}

\begin{document}

\begin{abstract}
\textbf{
Solid state spins in diamond, in particular negatively charged nitrogen-vacancy centers (NV), are leading contenders in the field of quantum sensing.
While addressing of single NVs offers nanoscale spatial resolution, many implementations benefit from using large ensembles to increase signal magnitude and therefore sensitivity. However, sensing with ensembles brings its own challenges given the random orientation of the spin quantization axis within the diamond crystal lattice. 
%The generalization from single NV to NV ensemble is not straightforward, especially in terms of electric and microwave fields coupling.
%The generalization from single NV to NV ensemble is not straightforward, especially in terms of coupling to the external field to be sensed and to the microwave field used for coherent manipulation and spectroscopy. %It affects all experimental parameters, most notably the coupling to the external field to be sensed and to the microwave field used for coherent manipulation and spectroscopy.
Here, we present an open source simulation tool that models the influence of arbitrary electric and magnetic fields on the electronic and nuclear spin states of NV ensembles, and can be extended to other color centers. Specifically, the code computes the transition strengths and predicts the sensitivity under shot-noise-limited optically-detected magnetic resonance. 
We illustrate the use of the code in the context of electric field sensing, a promising emerging functionality of NV centers with applications in biosensing and electronics, and bring several subtle features to light that are due to the interplay between different NV orientations and the external electric and microwave fields. %, recovering several published experimental results. 
Moreover, we show that our code can be used to optimize sensitivity in situations where usual arguments based on neglecting terms in the full Hamiltonian would give sub-optimal results. 
Finally, we propose a novel sensing scheme which allows to perform full vector electrometry without the need for precise bias magnetic field alignment, thus reducing the experimental complexity and speeding up the measurement procedure.}

\emph{keywords: nitrogen-vacancy (NV) centers, diamond, optically detected magnetic resonance (ODMR) spectroscopy, electric field sensing, numerical simulation, quantum metrology}

\end{abstract}

%\setboolean{displaycopyright}{false} %copyright statement should not display in the  supplementary document

%\maketitle

\section{Introduction}
\label{sec:introduction}

In recent years negatively charged nitrogen-vacancy (NV) centers in diamond, consisting of a nearest-neighbor pair of a substitutional nitrogen atom and a lattice vacancy (Fig.~\ref{fig:singleNV}a), have attracted a lot of attention for their long spin coherence times and favorable optical properties, making them promising candidates for quantum sensing and quantum information processing applications \cite{doherty2013nitrogen}. %Their long coherence time even at room temperature make them good candidates for quantum computation \cite{childress2013diamond}.  
Their nanoscale resolution, bio-compatibility, long coherence time even at room temperature and technical simplicity underpin their widespread use in quantum sensing -- notably for magnetic field, but also for electric field, temperature and strain sensing \cite{budker2007optical, schirhagl2014nitrogen, casola2018probing, radtke2019nanoscale, ho2022diamond, qiu2022nanoscale}.
NV centers sensing capabilities are based on optical preparation and readout of their spin states. The energy level structure depends on the NV center environment and can be probed through optically detected magnetic resonance (ODMR). The easiest approach is to continuously excite the sample, both optically (with green light) and with a coherent microwave field (MW), while collecting the photoluminescence (PL) signal (at red and near infrared wavelengths). In this work, we typically refer to this technique, named as continuous-wave (CW) ODMR. More advanced pulsed techniques can be used to improve the sensitivity and are reviewed for example in Ref~\cite{vandersypen2005nmr}. The spin transition spectra computed by our code can be used as input for further modeling in such contexts as well. 

An ODMR spectrum provides much information, such as the direction and the magnitude of a magnetic and electric fields at the NV center position. However, the presence of local intrinsic fields (e.g. due to strain), paramagnetic impurities, surface defects or unknown sample properties (e.g. random orientations of the NV centers), adds difficulties on the interpretation of the acquired ODMR spectrum \cite{jensen2013light}. Moreover, due to the interplay between the different quantities, the optimal sensing configuration is not always obvious.
The complexity is increased while using NV ensembles, due to the presence of all the 4 possible NV crystallographic orientations (each one with two possible NV vs. VN configuration) and to the higher concentration of other impurities compared to the single NV case in ultra-pure diamond substrate \cite{matsuzaki2016optically}. Even though a single NV center offers nanoscale spatial resolution, NV ensembles are a common solution in sensing applications since they allow to improve the signal-to-noise ratio (which ideally scales as $1/\sqrt{N}$,  $N$ being the number of NV centers involved), even if the sensitivity typically remains below the theoretical shot-noise limited sensitivity \cite{bauch2020decoherence, barry2020sensitivity}.

Here, we present a comprehensive sensing-oriented open source simulation tool that computes the ODMR spectrum from NV ensembles under arbitrary applied electric and magnetic fields. The two possible orientations for the nitrogen and the vacancy along a certain crystallographic direction are accounted for. %Moreover, a graphical interface is made available, to quickly plot the expected ODMR spectrum given a certain experimental configuration.
Analytical expressions are essential for understanding the physics of a system, but they are also typically obtained under simplifying hypotheses not always met in the experiment. Our simulation tool offers an easy way to interpret ODMR spectra: the user may choose to inspect one individual NV or all orientations together and can thereby conveniently assess new sensing schemes.
Special attention is dedicated to the CW-ODMR shot noise sensitivity under varying experimental settings (e.g. magnitude and direction of a bias magnetic field), thus helping the user to optimize the sensing scheme in a specific scenario.
%Besides the negative charge state we consider here, 
The simulation is based on $<100>$ oriented diamond containing ${}^{12}C$ and ${}^{14}N$ isotopes. Different situations (e.g. $<111>$ surfaces or implanted ${}^{15}N$) can easily be implemented by the user into the open-source code. Moreover, the code is most relevant under the condition of low-to-moderate magnetic field, $B < 100$ G. Our choice of approximations and their range of validity are described further in the paper.
%Indeed, at near-zero field randomly distributed charges and strain strongly impact the ODMR spectrum [REF] while at higher fields the ODMR contrast drops unless the magnetic field is perfectly align with one NV direction ~\cite{sharma2018imaging}. %[other assumptions we want to mention here? 13C? surface boundaries (focusing on ensemble the role of the surface can be typically neglected)?]

After providing the necessary theoretical background and describing the open-source code (Sec. \ref{sec:methods}), we present some application examples in the context of electric field sensing (Sec. \ref{sec:esensing}-\ref{sec:2nvorientations}). In particular, the impact of both the direction and magnitude of a bias magnetic field on the electric field sensitivity is discussed. This information is essential for the optimal design of an experiment or a sensing device and for the correct interpretation of experimental results. 
Moreover, we present a novel electric-field sensing scheme with NV ensemble that is based on the alignment of a bias magnetic field orthogonal to two NV families and allows to retrieve the whole vector information. A quantitative study of this new configuration would not be possible without solving the full Hamiltonian as our code is doing. 

%Considering the interaction between the dipole moment of the spin transitions and the MW probing field \cite{kolbl2019determination, herrmann2016polarization} (i.e. polarization and frequency), not only the resonant frequencies are calculated, but also the corresponding transition amplitudes, from which the ODMR spectrum can be easily obtained. [should we discuss more this point? is it precise enough?] %The two possible orders of nitrogen and vacancy along a certain crystallographic direction is included.

%%%%%%%%%%%%%%%%%%%%%%%%%%%%%%%%%%%%%%%%%%%%%%%%%%%%%%%%%%%%%%%%%%%%%%%%%%%%%%%%%%%%%%%%%%%%%%%%%%%%%%%%%%%%%%%%%%%%%%%%%%%%%%%%%%%%%%%%%%%%%%%%%%%%%%%%%%%%%%%%%%%%%%%%%%%%%%%%%%%%%%%%%%%%%%%%

\section{Methods}\label{sec:methods}
%[here I would put a short introduction about the code itself, like programming language, structure, and any other related informative thing]
The code we developed is written in Python, and can be accessed and downloaded here \url{https://github.com/chris-galland/NV-ODMR-simulation}. %This can be useful for sharing and improving it.
In this section we provide the necessary theoretical background and detail its main features. More technical details can be found directly in the code.
Note that we did not aim to include all possible physical interactions and variations of configurations (e.g. nitrogen and carbon isotopes, diamond cut angle, etc.) into the simulation; however, the interested users can further extend its functionalities, based on their specific requirements. %For example, the effect of $^{15}N$ or the impact of strain in the diamond lattice could be included.

\subsection {From single NV Hamiltonian to ODMR spectrum}\label{sec:singleNV}

\begin{figure}
\centering
\includegraphics[width=0.8\columnwidth]{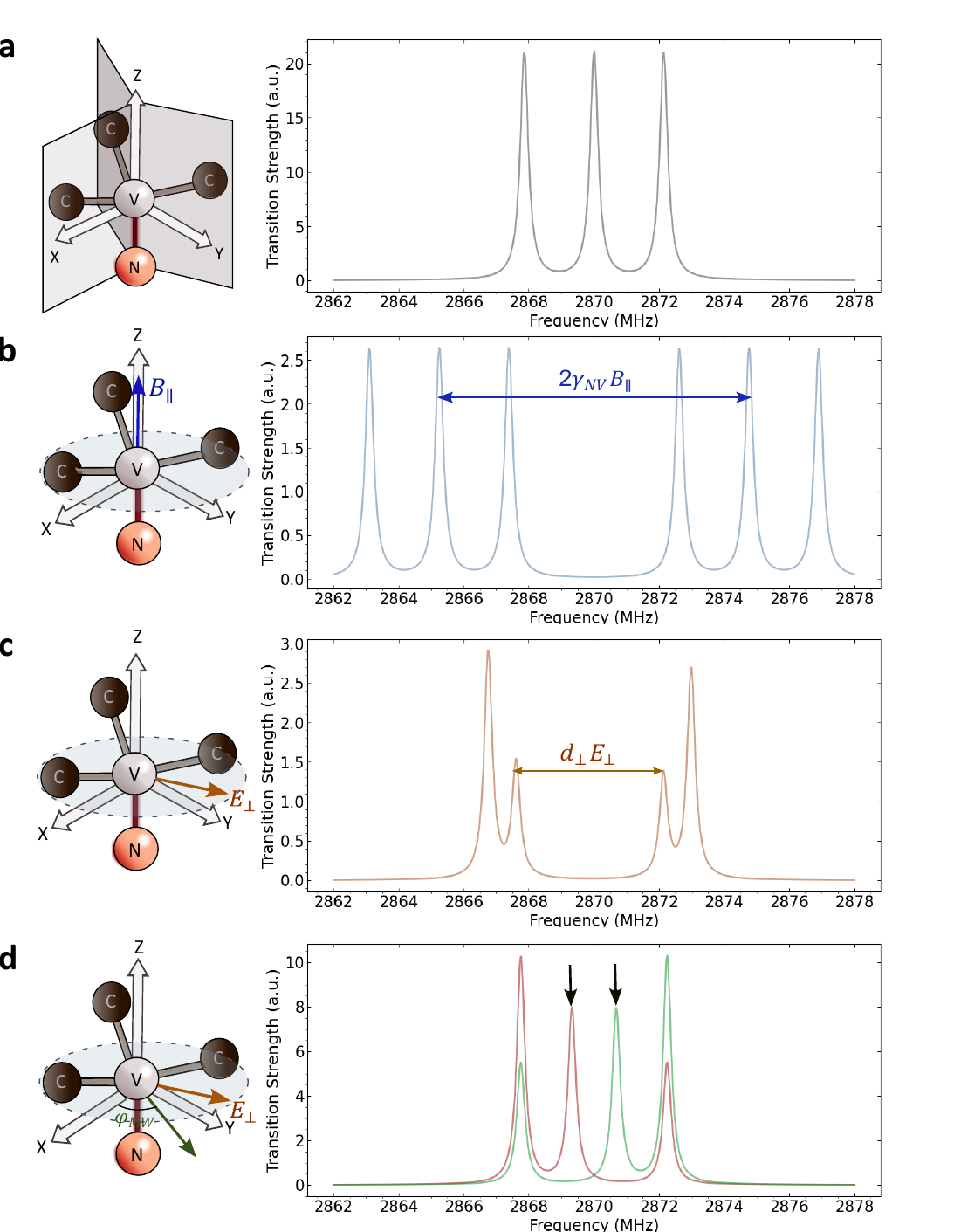}
\vspace*{0cm}
\caption{(a) Left panel: Atomic structure of the NV center and its three reflection planes, with the chosen coordinate system $(x,y,z)$ . Right panel: Computed spin transition strength spectrum without any applied static field nor strain, at room temperature, assuming an electronic spin transition linewidth $\delta = 0.3$ MHz (corresponding to $T_2^* = 1.1~\mathrm{\mu s}$). The hyperfine structure due to the ${}^{14}N$ nuclear spin is clearly visible. (b) Transition strength spectrum in the presence of a static magnetic field ($B=1.8$~G) aligned along $z$. (c) Transition strength spectrum in the presence of an electric field ($E_{\bot}=4\cdot10^7$~V/m) in the $xy$ plane (no magnetic field applied). 
(d) Transition strength spectrum for two different MW linear polarisations lying in the $xy$ plane ($\phi_{MW}=0, \pi/2$ with respect to $x$). An electric field ($E_{\bot}=4\cdot10^6$~V/m) with the same direction as in (c) is applied in this case.
%NV spin interacts with the polarized microwave field on its x-y plane: MW with 3*pi/4(coca), pi/4(grey) defined in the NV frame
}
\label{fig:singleNV}
\end{figure}

%\subsubsection{H for single NV and single NV coordinate}
Given an NV center, we define the NV reference frame by a direct orthonormal coordinate system $(x,y,z)$, where $z$ is along the NV major symmetry axis (also simply referred to as NV axis) and points from the nitrogen atom to the vacancy. The $x$ axis belongs to one of the defect's three symmetry planes containing a carbon atom neighbouring the vacancy, see Fig.~\ref{fig:singleNV}a. In the following, we will decompose the fields of interest into parallel ($\parallel$) and transverse ($\bot$) components with respect to the NV axis.
The ground state spin Hamiltonian of an NV center can be written as~\cite{doherty2013nitrogen,gali2019ab}:
\begin{equation}
\label{eq:sumH}
    \hat{H}_{tot} = \hat{H}_{el} + \hat{H}_{nuc}
\end{equation}
where $\hat{H}_{el}$ refers to the electronic spin Hamiltonian and $\hat{H}_{nuc}$ to the hyperfine interaction between the NV electron spin and the nitrogen nuclear spin.
In the natural electronic spin-triplet basis $\{|m_s=-1\rangle,|m_s=0\rangle,|m_s=1\rangle \}$ and in the presence of an external magnetic field $\vec{B}$ and electric field $\vec{E}$, $\hat{H}_{el}$ can be written as:
\begin{equation}
	\label{eq:Hel}
	\frac{1}{h} \hat{H}_{el} = (D_{gs} + d_{\parallel} E_{z} ) \left({\hat{S}_z}^2 - \frac{2}{3} \right) + \gamma_{NV} \vec{B} \cdot \hat{\vec{S}} - d_{\bot} E_{x} ({\hat{S}_{x}}^2 - {\hat{S}_{y}}^2) + d_{\bot } E_{y} (\hat{S}_{x} \hat{S}_{y} + \hat{S}_{y} \hat{S}_{x})
\end{equation}
where $D_{gs}$ is the temperature-dependent zero field splitting ($D_{gs}=2.870 ~\mathrm{GHz}$ at room temperature), $\gamma_{NV}  = \frac{\mu_{B}}{h} \cdot g_{NV} = 2.80 ~ \mathrm{MHz/G}$ is the gyromagnetic ratio, $h$ is the Planck constant, $\mu_{B}$ is the Bohr magneton. %Note that working in the regime of relatively weak magnetic fields, we assume the ground state electron g-factor to be isotropic.
Additionally, $d_{\parallel} = 0.35 \mathrm{~Hz~cm V}^{-1}$ and $d_{\bot } =  17 \mathrm{~Hz~cm V}^{-1}$ are the electric-field coupling strengths for parallel and transverse components. Note that $d_{\bot }$ is about 50 times larger than $d_{\parallel}$, offering correspondingly larger sensitivity to the electric field component normal to the NV axis.
$\hat{\vec{S}} = ( \hat{S}_x, \hat{S}_y, \hat{S}_z)$ is the total electronic spin operator; since for NV centers the total spin is $S=1$, the three operators can be expressed in matrix form as: 
\begin{equation}
\hat{S}_x = \frac{1}{\sqrt{2}} \begin{pmatrix} 0&1&0\\1&0&1\\0&1&0 \end{pmatrix} ;
\hat{S}_y = \frac{1}{\sqrt{2}} \begin{pmatrix} 0&-i&0\\i&0&-i\\0&i&0 \end{pmatrix} ;
\hat{S}_z = \frac{1}{\sqrt{2}} \begin{pmatrix} 1&0&0\\0&0&0\\0&0&-1 \end{pmatrix}
\label{eq:pauli}
\end{equation}
Strain in the diamond lattice can also be accounted for in $\hat{H}_{el}$ in the form of an effective electric field~\cite{udvarhelyi2018spin, barson2017nanomechanical}. 

In the case of a ${}^{14}N$ nucleus, the contribution $\hat{H}_{nuc}$ to the hyperfine electronic level structure can be written as:
\begin{equation}
	\label{eq:Hnuc}
	\frac{1}{h} \hat{H}_{nuc}  = A_{\parallel}\hat{S}_{z}\hat{I}_{z} + A_{\bot } (\hat{S}_{x} \hat{I}_{x} + \hat{S}_{y} \hat{I}_{y}) + P \left(\hat{I}_z^2-\frac{2}{3}\right) + \gamma_{N} \vec{B} \cdot \hat{\vec{I}}
\end{equation}
where $A_{\parallel}= -2.14~\mathrm{MHz}$ and $A_{\bot }= -2.7~\mathrm{MHz}$ are the ${}^{14}N$ axial and transverse magnetic hyperfine parameters respectively, $P = -4.95~\mathrm{MHz}~$ is the nuclear quadrupole parameter and $\gamma_N = \frac{\mu_{B}}{h} \cdot g_{N} = 0.31 ~ \mathrm{kHz/G}$ is the nuclear gyromagnetic ratio ($g_{N} =  0.404$ is the nuclear $g$-factor and $\mu_{N}$ the nuclear magneton). $\hat{\vec{I}} = ( \hat{I}_x, \hat{I}_y, \hat{I}_z)$ is the nuclear spin operator of ${}^{14}N$. Since the total nuclear spin is $I = 1$,  the Pauli matrices of the cartesian operators in the natural nuclear spin basis $\{ |m_I=-1\rangle,|m_I=0\rangle,|m_I=1\rangle \}$ are exactly the same as reported in Eq.~\ref{eq:pauli}. The extension to ${}^{15}N$ nucleus, which has spin I = $\frac{1}{2}$, could be easily implemented by replacing the matrix for nuclear spin operator and deleting the nuclear quadrupole interaction term since only nuclei with spin $I\geq 1$ may possess electric quadrupole moments.

In the code, the computation of the electronic energy levels in presence of external fields is straightforward and consists in diagonalizing the complete Hamiltonian in Eq.~\ref{eq:sumH} and extracting its eigenvalues. The corresponding eigenvectors are represented in the natural spin basis states: ${|m_s, m_I\rangle}$, where $m_s=0,\pm1$ and $m_I=0,\pm1$ refer to the $z-$projection of the electronic and nuclear spin respectively.

%After diagonalizing the sum of the above mentioned Hamiltonian, the obtained eigenvalues of the $ \hat{S} $ , $\hat{I}$ operator for an individual NV center are $m_s = 0,\pm 1$ and $m_I = 0,\pm 1$ correspondingly. Therefore, this simulated quantum system has three possible electronic states and three possible nuclear state. In total, $3 \times3 $ states which corresponds to nine energy levels will represent the current quantum state $| m_s, m_I \rangle$ of this quantum sensor. 

%\subsubsection{Magnetic dipole transitions and the MW  driving field --- do we want to keep this title? and in general the structure of title A and subtitles?}Yuchun: I include the subtitle just formyself to understand the structure
In CW-ODMR, the spin transitions are probed by sweeping the applied MW field frequency that interacts with the magnetic dipole moment of the NV center~\cite{dobrovitski2013quantum,doherty2012theory}, see Fig.~\ref{fig:singleNV}d.
In order to reproduce the CW-ODMR spectrum, our code models the coupling of a linearly polarized MW magnetic field $\vec{B}_{MW} = (B^{MW}_x,B^{MW}_y,B^{MW}_z)$ to the electronic and nuclear spin through the following interaction Hamiltonian, denoted as $\hat{H}_{int}$ :
%# Total interation Hamiltonian
\begin{equation}
\label{eq:gshamiltonian}
   \frac{1}{h} \hat{H}_{int} = \gamma_{NV} \vec{B}_{MW} \cdot  \hat{\vec{S}} + \gamma_{N} \vec{B}_{MW} \cdot \hat{\vec{I}}
\end{equation}
Note that arbitrary MW polarization can be implemented into the code by considering a complex MW magnetic field vector $\vec{B}_{MW}$~\cite{kolbl2019determination}.
The above interaction Hamiltonian induces transitions between an initial eigenstate $|i \rangle$ of energy $E_i=h  \nu_{i}$ and a final eigenstate $|f \rangle$ of energy $E_f=h \nu_{f}$. The magnetic dipole transition probability between these states is calculated as: 
\begin{equation}
    T_{i,f} \propto | \langle f| \hat{H}_{int}|i\rangle|^2
\end{equation}
Within the program, we assume that the transition between $|i\rangle$ and $|f\rangle$ has a Lorentzian shape, with amplitude given by $T_{i, f}$ and linewidth $\delta$ \cite{kolbl2019determination}. This linewidth depends on the specific experimental conditions (diamond growth, laser and MW power, etc.) and is given to the code as an input parameter. The frequency-dependent transition amplitude between the state $|i\rangle$ and the state $|f\rangle$, induced by the MW with frequency $\nu$ is thus:

\begin{equation}
\label{eq:lore}
    T_{i, f}(\nu) = \frac{T_{i, f} {\delta ^2}}{4 \left[ \left(\nu - |\nu_{f}-\nu_{i}| \right)^2 + \frac{{\delta ^2}}{4}\right]}
\end{equation}

%When the swept MW frequency is resonant with the energy difference between final state and initial state, namely $\nu_{MW} - |\nu_{f}-\nu_{i}|^2 = 0 $, the transition strength is maximum $T_s = T_A$. On the contrary, when $|\nu_{MW} - |\nu_{f}-\nu_{i}|^2| \gg 0 $, $T_s \sim 0 $. 

Eventually, the total transition strength for a given MW frequency is obtained by summing over all possible initial and final states: $T(\nu) = \sum_{i, f} T_{i, f}(\nu)$. 
Since $m_s=\pm 1$ and $m_s=0$ exhibit different PL intensities, the transition strength $T(\nu)$ is proportional to the ODMR contrast. 
Note that this is true only in the low magnetic field regime ($B < 100$~G), which our study is limited to. Higher magnetic fields induce mixing of the excited states and therefore we cannot assume anymore that $T$ is proportional to the ODMR contrast. In this case, a model including decay and pumping rates between each ground, excited and metastable state is needed to reproduce the PL and ODMR response~\cite{tetienne2012magnetic}.

To conclude this section, we briefly summarize the influence of external fields on the transition strength spectrum. Examples of the final computed spectra in different conditions that illustrate these influences are shown in Fig.~\ref{fig:singleNV}. The ground state energy level structure without any perturbation is shown in Fig.~\ref{fig:singleNV}a. It is dominated by zero field splitting ($D_{gs}=2.87$ GHz) and the hyperfine interaction due to the ${}^{14}N$ nucleus that breaks the degeneracy of the $m_I=0,\pm1$ states, leading to a characteristic triplet of peaks.
%Each level is still degenerate ($|m_s=1\rangle$).  

A magnetic field acts on the level structure through the Zeeman effect. In particular, in presence of an axial $\vec{B}$ field, the Zeeman splitting is proportional to the axial component of the field (Fig.~\ref{fig:singleNV}b), which is the basis for most magnetic sensing protocols developed in the recent years \cite{casola2018probing,barry2020sensitivity,rondin2014magnetometry}. %On the contrary, the hamiltonian will not be diagonal if a transverse field is present. this means that the eigen vector have no definite spin.

Experimentally, the Stark shift induced by an electric field is more difficult to observe due to the smaller coupling constants \cite{dolde2011electric, michl2019robust}.
An axial electric field shifts all $|m_s=\pm1\rangle$ states with respect to the state $|m_s=0\rangle$ by $d_\parallel E_{z}$. On the other hand, a transverse electric field makes the electronic Hamiltonian $H_{el}$ non-diagonal, which results in mixing of states $| m_s = \pm1 \rangle$. In the absence of magnetic field, the new eigenvectors are: 
\begin{equation}
\begin{aligned}
    |-\rangle = \frac{1}{\sqrt{2}}( e^{i \phi_{E}}|m_s =+1\rangle+|m_s=-1\rangle)\\
    |+\rangle = \frac{1}{\sqrt{2}}( e^{i \phi_{E}}|m_s=+1\rangle-|m_s=-1\rangle)
\end{aligned}
\end{equation}
%with $\phi_{\Pi}=\arctan(\Pi_x/\Pi_y)$.
with $\phi_{E}=\arctan(E_x/E_y)$. Correspondingly, the energy levels present a splitting in the central resonance proportional to $E_{\bot}$, as illustrated in Fig.~\ref{fig:singleNV}c.
%Since the states $|\pm \rangle$ are a coherent superposition of the states $|m_s=\pm 1 \rangle$, the ODMR resonance is observed also in this case as a reduction in the fluorescence emission at the new MW resonance frequencies. [this is currently almost copied and pasted from \cite{petrini2020quantum}, but I like this clarification] %find a place later, no ODMR discussion for the moment
Stark shift has been used to image electric fields~\cite{dolde2011electric, michl2019robust, qiu2022nanoscale, broadway2018spatial} as well as local strain and stress~\cite{trusheim2016wide, kehayias2019imaging, barfuss2019spin}, both with single NV and ensembles. Noteworthy, in diamond with high NV concentration, a central splitting can be observed even in absence of external electric fields, due to local electric fields caused by surrounding NV centers and other spin impurities \cite{mittiga2018imaging} as well as lattice strain \cite{levchenko2015inhomogeneous}.
%Typically this central feature can be easily observed in diamond with high NV concentration, even in absence of external electric fields, due to local electric fields caused by surrounding NV centers and other spin impurities \cite{mittiga2018imaging} as well as lattice strain \cite{barfuss2019spin}. %[Just on thing: in the second paper cited here they precisely make the opposite point: that the splitting is NOT due to strain, but to local electric field due to neighboring impurities (P1 or other NV centers) - I corrected].
The combined influence of magnetic and electric field is more complex, and is discussed in Sec.~\ref{sec:esensing}.

%missing the discussion of MW polarization on hyperfine structure....ref.\ref{fig:singleNV}(d). Good point, I think we need to add something about it. Here it seems a great place where do it.

%Meanwhile, unpolarized MW will increase the complexity of the obtained hyperfine transitions. [No]

Finally, the linear MW polarization has an influence on the amplitudes of allowed transitions. In particular, in the presence of an effective electric field only, an analytical expression for these amplitudes can be derived. For a MW magnetic field orthogonal to the NV axis and with polar angle $\phi_{MW}$, the transition amplitude between $|m_s=0\rangle$ and $|\pm\rangle$ eigenstates introduced above is given by~\cite{mittiga2018imaging, kolbl2019determination}:
\begin{equation}
	\label{eq:imbalance}
	T_{0, \pm} \propto (1 - \cos(2\phi_{MW} + \phi_E))
\end{equation}
This is illustrated in Fig.~\ref{fig:singleNV}d, where for $m_I=0$ states (inner hyperfine states, indicated by black arrows) the transition amplitude at lower frequency is completely suppressed for $\phi_{MW}= 0$ and the upper frequency transition has maximal amplitude, whereas it is the opposite for $\phi_{MW}=\pi/2$. Note that the proportionality factor omitted in Eq.~\ref{eq:imbalance} depends on the transverse component of the MW field (transition amplitude is maximized for completely transverse $\vec{B}_{MW}$) and is different for the different nuclear spin states $m_I$. The dependence on linear polarization has been used to reconstruct local effective electric fields (with possible contributions from strain) for single NV centers\cite{mittiga2018imaging, kolbl2019determination}.

The presence of a magnetic field also contributes to the MW polarization dependence, in particular adding a phase that depends on $\phi_B$ to the cosine term in Eq.~\ref{eq:imbalance}~\cite{kolbl2019determination}. Deriving an analytical expression for the combined influence of electric and magnetic field on MW polarization response is beyond the scope of our discussion. However, this aspect can be explored numerically using our code.

\subsection {From single NV to NV ensembles}
 
%\subsubsection{Coordinate transformation}
%\begin{figure}
%\centering
%\includegraphics[width=0.9\textwidth]{logos/figure2option3.pdf}
%\vspace*{0cm}

%\label{fig:coordinate}
%\end{figure}

\begin{figure}
\centering
\includegraphics[width=0.9\textwidth]{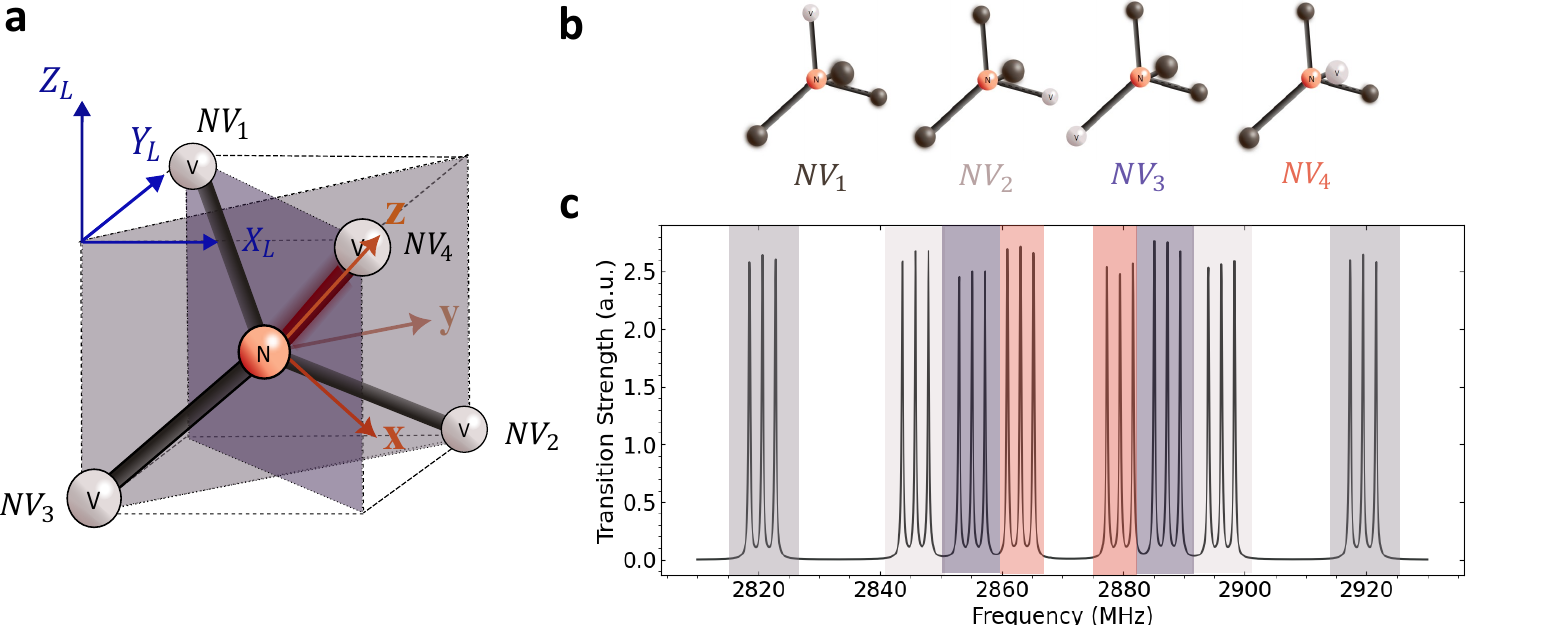}
\vspace*{0cm}
\caption{ (a) NV reference frame $(x,y,z)$ vs. lab frame $(X_L,Y_L,Z_L)$ for the most usual (100) oriented diamond crystals; $z$ points along the N-to-V direction. (b) Schematic of the four possible NV orientations inside the diamond crystal lattice. (c) Transition strength spectrum in the presence of a static magnetic field ($B=18$~G) whose direction is chosen such that all resonances are clearly resolved. There are $3 \times 8$ peaks corresponding to the four different NV orientations: each giving rise to two non-degenerate electronic spin transitions ($m_s=0$ to $\pm 1$) and each with three hyperfine transitions.
}
\label{fig:coordinate}
\end{figure}

The ground state Hamiltonian for a single NV center, Eq.~\ref{eq:sumH}, is defined in the NV reference frame. However, we typically deal with external fields whose direction is expressed in the laboratory frame $(X_L,Y_L,Z_L)$, see Fig.~\ref{fig:coordinate}a. For ease of use, our code employs polar rather than cartesian coordinates to express external fields. 

When dealing with NV ensembles, each NV center is aligned along one of the 4 possible crystallographic axes (Fig. \ref{fig:coordinate}b, upper panel). The four possible NV orientations, named as $NV_i, i=1,2,3,4$, are typically equally present (and each one presents two possible directions, NV and VN); exceptions are under particular growth conditions leading to preferential orientation along one axis \cite{michl2014perfect,pham2012enhanced,edmonds2012production}. In the code, we assume the NV centers equally distributed among the 4 possible crystallographic axes, but different situations can be easily implemented by weighting the contributions accordingly.
%Moreover, for each defect axis, two equally likely configurations are taken into account, depending on the respective  position of nitrogen vs vacancy along the defect axis.
In order to compute the energy levels and the corresponding transition strength spectra for an NV ensemble, we proceed in three main steps:

1. The input fields ($\vec{E}$ and $\vec{B}$), given in polar coordinates in the laboratory frame, are converted into the different $NV_i$ frames. %NV_2, NV_3, NV_4$). 
The same holds for the MW field used to probe the spin transitions. %In particular, the component parallel and transverse to each NV axis is computed. 
The relation between the laboratory and the $NV_i$ reference systems depends on the diamond surface considered. In our case we assume the most commonly used $<100>$ surface. Different cases can be implemented changing the transformation matrices accordingly (or applying an additional global transformation). %See Fig. \ref{fig:coordinate}(a).
%For analysis of those external perturbations, a separate coordinate system for each NV alignment from that in the lab frame should be utilized.

2. The Hamiltonian in each $NV_i$ frame is numerically diagonalized and the corresponding transition strength spectrum obtained using the procedure explained for the single NV case (See Sec. \ref{sec:singleNV} ).

3. The transition strength spectra corresponding to the different $NV_i$ frames are summed together, obtaining the complete spectrum.

%Meanwhile, differing from the method implemented in \cite{ye2019reconstruction}, where the projection along $\hat{z}$ axes for one NV frames is computed and corresponding four polar angles with respect to four different NV axes are obtained. The static vector fields in our case are projected onto each of the NV axes $ ( \hat{e}_{NV1} , \hat{e}_{NV2} , \hat{e}_{NV3} , \hat{e}_{NV4} )$ corresponds to four distinct NV reference frames. The advantage of this step is that the relative phase between external $\vec{B}$ and $\vec{E}$ for each NV axes could be known, since the value of this phase between the magnetic field and electric field is essential to observe the splitting from the resulting electric field (add a refernece).

%The next step is to obtain 4 transformation matrices for those 4 NV frames by using geometric arguments to transform the tetrahedral components $( \hat{x}^{'}, \hat{y}^{'}, \hat{z}^{'})$ into a Cartesian coordinate $( \hat{x}, \hat{y}, \hat{z})$. 
%For example, in the case of constructing the NV frame for $\hat{e}_{NV4}$ (indicated by blue arrows), the relative orientation between the NV4 frame and the triaxial orthogonal frame is shown schematically in \ref{fig:coordinate}(a). 

Referring to Fig. \ref{fig:coordinate}a, the transformation matrices between the laboratory frame and the different $NV_i$ frames are given by:
%Thus the NV reference frame is defined via the following three orthonormal basis vectors: \[ 
% \hat{x}^{'} = \frac{1}{\sqrt{6}} \begin{pmatrix} 1\\-1\\-2 \end{pmatrix} 
% \hat{y}^{'} =  \frac{1}{\sqrt{2}} \begin{pmatrix} 1\\1\\0 \end{pmatrix}
% \hat{z}^{'} = \hat{e_{NV1}} = \frac{1}{\sqrt{3}} \begin{pmatrix} 1\\-1\\1 \end{pmatrix}
%\]

\begin{equation*}
  T_{NV_1} = 
    \begin{pmatrix}
    -1/\sqrt{6} & 1/\sqrt{6} & -2/\sqrt{6}\\
    -1/\sqrt{2} & -1/\sqrt{2} & 0\\
    -1/\sqrt{3}& 1/\sqrt{3} & 1/\sqrt{3}
    \end{pmatrix} 
    T_{NV_2} = 
    \begin{pmatrix}
    1/\sqrt{6} & 1/\sqrt{6} & 2/\sqrt{6}\\
    1/\sqrt{2} & -1/\sqrt{2} & 0\\
    1/\sqrt{3}& 1/\sqrt{3} & -1/\sqrt{3}
    \end{pmatrix}
\end{equation*}

\begin{equation*}
    T_{NV_3} = 
    \begin{pmatrix}
    -1/\sqrt{6} & -1/\sqrt{6} & 2/\sqrt{6}\\
    -1/\sqrt{2} & 1/\sqrt{2} & 0\\
    -1/\sqrt{3}& -1/\sqrt{3} & -1/\sqrt{3}
    \end{pmatrix}
    T_{NV_4} = 
    \begin{pmatrix}
    1/\sqrt{6} & -1/\sqrt{6} & -2/\sqrt{6}\\
    1/\sqrt{2} & 1/\sqrt{2} & 0\\
    1/\sqrt{3}& -1/\sqrt{3} & 1/\sqrt{3}
    \end{pmatrix}
\end{equation*}

\begin{figure}
\centering
\includegraphics[width=0.7\columnwidth]{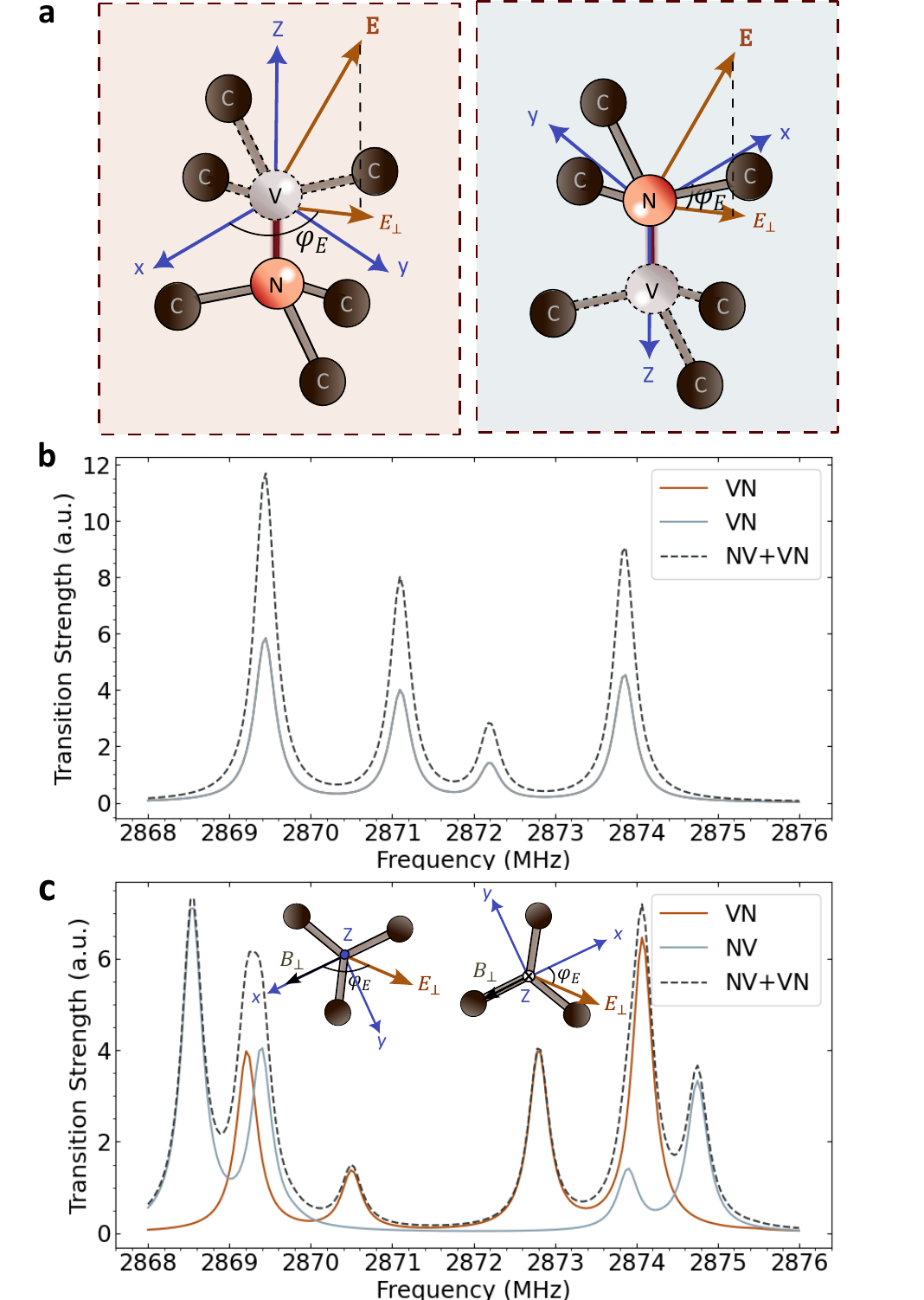}
\vspace*{0cm}
\caption{(a) Two possible configurations for a given NV axis with their respective reference frames. (b) Transition strength spectrum containing equal contributions from both NV and VN under an applied static magnetic field ($B_\perp=20$~G) perpendicular to the NV axis. The solid curves corresponding to NV and VN are perfectly overlapping. The dashed curve indicates the summation of both contributions. (c) Transition strength spectrum of both NV and VN defects under an applied combination of static electric field ($E_{bot}=1\cdot10^7$~V/m) and static magnetic field ($B_\perp=20$~G) perpendicular to the NV axis. The projection of NV and VN with 3 carbon atoms next to the nitrogen are plotted. In the NV frame the applied electric field makes an angle $\phi_E$ with respect to $x$ axis resulting in frequency upshift (orange line). The same electric field vector in the VN frame has a different azimuthal angle resulting in frequency  downshift (blue line).}
\label{fig:nvvn}
\end{figure}

Further, as illustrated in Fig. \ref{fig:nvvn}a, given a certain crystallographic axis there are two possible arrangements for the nitrogen atom and the vacancy, that we refer to as NV and VN. In typical growth conditions we can assume these two configurations to be equally present. 
In their respective reference frame, as defined in Sec. \ref{sec:singleNV}, VN and NV centers experience opposite fields (i.e. for a magnetic field: $\vec{B}_{VN} = -\vec{B}_{NV}$). A closer look at Eq.~\ref{eq:Hel} reveals that the Hamiltonian remains identical upon a simultaneous change $\vec{B}\rightarrow-\vec{B}$ and $m_s = \pm1 \rightarrow \mp1$. Thus, NV and VN respond in an identical way to an external magnetic field. A similar argument can be made for a longitudinal electric field. However, a transverse electric field breaks this inversion symmetry and has a different contribution on the two configurations. 

This fact can be appreciated in Fig.~\ref{fig:nvvn}b and c, where we report the transition strength spectrum for NV and VN configurations in the presence of external fields. Fig.~\ref{fig:nvvn}b corresponds to the spectrum in presence of a transverse magnetic field ($B_\perp = 20$~G): in this case the two curves perfectly overlap, confirming that NV and VN response is identical. Fig.~\ref{fig:nvvn}c corresponds to the spectrum in presence of an electric field transverse to the $NV$ axis (a bias static magnetic field $B_\perp=20$~G is applied in the same direction):  the two configurations show different splitting widths, confirming that considering both NV and VN is essential\cite{yang2020vector}.
To fully model NV ensemble within our code, we thus compute the transition strength spectrum for both NV and VN arrangements, and sum them to obtain the final spectrum.
The full transition strength spectrum per defect center $T(\nu)$ is obtained by summing the contribution of NV and VN for all 4 NV orientations. To correctly model the contribution of each NV axis and direction to the PL, we divide by the total number of configurations (i.e 8). The resulting spectrum is illustrated in Fig.~\ref{fig:coordinate}c in the presence of an external magnetic field. The direction of magnetic field was chosen in order to allow each NV orientation to be resolved.

\section{Optimization of electric field sensing}\label{sec:esensing}

\begin{figure}
	\centering
	\includegraphics[width=\columnwidth]{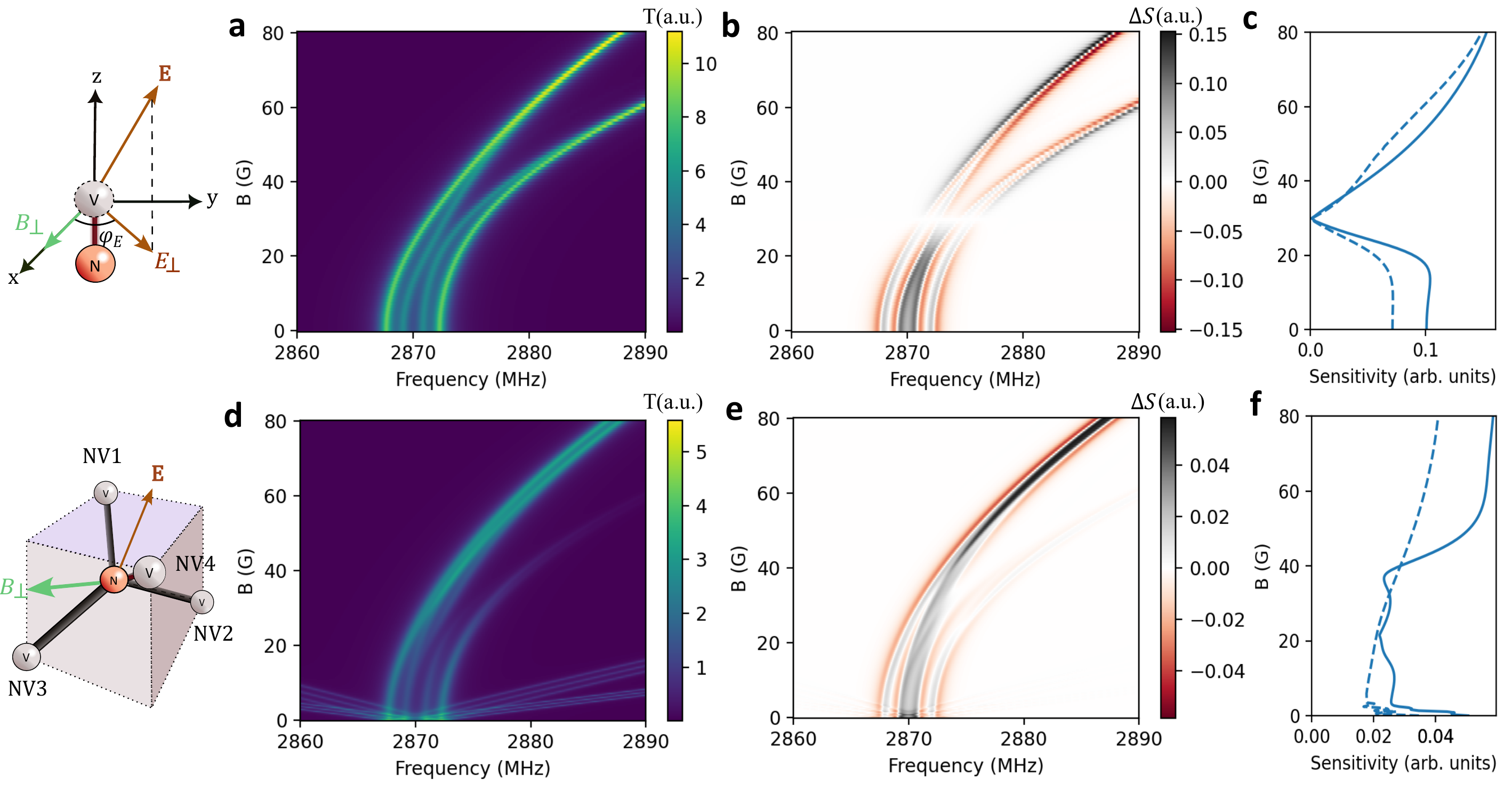}
	\caption{Electric field sensitivity optimization with (a-c) a single NV and (d-f) an ensemble. (a) Transition strength spectrum versus static magnetic field amplitude $B$. (b) Electric field sensitivity $\Delta S$ versus $B$. (c) Optimal values of the sensitivity (maximum: solid line; minimum (absolute value): dashed line) of $\Delta S$, over MW frequency $\omega$, for each $B$ value. $\vec{B}, \vec{E}$ are orthogonal to the NV axis, with $\phi_B = 0, \phi_E = \pi/4$ and $E = 5\cdot10^6~\mathrm{V/m}$. $\delta = 1~\mathrm{MHz}$. MW is unpolarized, with MW field $\vec{B}_{MW}$ orthogonal to NV axis. $\Delta E = 1\cdot10^5~\mathrm{V/m}. $ (d-f) Same plots for an NV ensemble when the MW field is linearly polarized with $\vec{B}_{MW}$ orthogonal to $NV_1$ axis, and $\phi_{MW} = \pi/2$ in $NV_1$ frame. Values for $\vec{B}, \vec{E}$ are unchanged. }
	\label{fig:Efield_opti}
\end{figure}

In this section, we illustrate how our code can be used to optimize the sensitivity of electric field measurements using an NV ensemble.
We start with a short general discussion on electric field sensing with a single NV center. For the sake of simplicity, we neglect the effect of strain and local stray electric fields.
The standard sensing scheme for electric fields relies on a bias magnetic field applied orthogonal to the NV axis, typically of a few tens of Gauss\cite{dolde2011electric, li2020nanoscale, michl2019robust}. In this way, the Zeeman shift is suppressed and the eigenstates become sensitive to electric fields while being decoupled from longitudinal magnetic noise. 
The effect of electric field is maximal on states with $m_I = 0$. In the presence of both an electric field $\vec{E}$ and a transverse magnetic field $\vec{B}_\perp$, the transition frequencies for this sub-ensemble of states are~\cite{dolde2011electric, doherty2012theory}:
\begin{equation}
	f_\pm\left(\vec{E}, \vec{B}_\perp \right) = D_{gs} + d_\parallel E_z + 2\Lambda \pm \sqrt{d_\perp^2 E_\perp^2 - 2 \Lambda d_\perp E_\perp \cos{\phi} + \Lambda^2}
	\label{eq:transition_freq}
\end{equation}
where $E_z$ and $E_\perp$ are the electric field components respectively longitudinal and transverse to the NV axis, $\Lambda = \gamma_{NV}^2 B_\perp^2 / 2 D_{gs}$ and $\phi = 2\phi_B + \phi_E$, with $\phi_{B, E}$ the polar angle of $\vec{B}_\perp$ and $\vec{E}$ in the transverse plane (see Fig.\ref{fig:Efield_opti}a). Note that for states with $m_I=\pm1$, the hyperfine interaction acts as a small longitudinal magnetic field, hence reducing the sensitivity to electric fields.
Taking into account that $d_\parallel \ll d_\perp$, the influence of longitudinal electric field is much weaker than transverse ones, and in realistic experiments precise sensing of $E_z$ is not possible. In the following we ignore $E_z$ and focus on transverse electric fields.

We distinguish three regimes of electric fields: (i) weak fields, with $d_\perp E \ll \Lambda, \left|A_\parallel\right|$ (typically $E \lesssim 10^5~\mathrm{V/m}$) ; (ii) moderate fields ($10^5~\mathrm{V/m} \lesssim E < 5 \cdot 10^7~\mathrm{V/m}$) ; and (iii) strong fields, $d_\perp E > \left|A_\parallel\right|, \Lambda$ ($E > 5 \cdot 10^7~\mathrm{V/m}$). 

Even though case (i) requires the best sensitivity and is the most explored case with single NV~\cite{dolde2011electric, doherty2014measuring, barson2021nanoscale, qiu2022nanoscale}, cases (ii) and (iii) are still relevant in various contexts~\cite{broadway2018spatial, michl2019robust, yang2020vector}. In particular, $E$ field sensing with NV ensemble may be limited to moderate and strong fields due to the presence of local electric fields and inhomogenous strain that might screen weak E fields.

In case (i), only states with $m_I=0$ are affected by $\vec{E}$. Moreover, Eq.~\ref{eq:transition_freq} can be approximated as $f_\pm(\vec{E}_\perp, \vec{B}_\perp) \approx D_{gs} + (2 \pm 1) \Lambda \mp d_\perp E_\perp \cos{\phi}$, from which it appears that the electric field-induced shift of the transitions is independent of $| \vec{B}_\perp|$, but is maximal for $\phi = 2\phi_B + \phi_E = 0$ or $\pi$. As a consequence, the $B$ field direction defines the axis of optimal sensing. 

%Case (iii) can also be relevant in specific contexts~\cite. Here, since $d_\perp E > \left|A_\parallel\right|$, states with $m_I=0$ become degenerate with the outer hyperfine states.  

In case (ii), Eq.~\ref{eq:transition_freq} can still be used to compute the transitions frequencies for $m_I=0$ states. However, other hyperfine transitions can no longer be ignored, and optimal sensitivity conditions might deviate from single-transition analytical predictions from Eq.~\ref{eq:transition_freq}. This case is illustrated in Fig.~\ref{fig:Efield_opti}a, which presents the transition amplitudes for a single NV center versus transverse magnetic field, for $E = 5 \cdot 10^6~\mathrm{V/m}$, $\phi_E = \pi/4$ and $\phi_B = 0$. We notice that for $B>40$~G the inner and outer transitions begin to merge due to their non-zero linewidth. For simplicity, we consider unpolarized microwave excitation for the moment.

In order to estimate the sensitivity to electric field for a given $B$ value, we compute the differential spectrum $\Delta S(\nu) = \Delta T(\nu) / \Delta E$, with $\Delta T(\nu) = T(\nu, \vec{E} + \Delta\vec{E}, \vec{B}) - T(\nu, \vec{E}, \vec{B})$ and $\Delta E \ll E$. $\Delta S$ is proportional to the variation of photoluminescence intensity caused by a change of electric field, thus it is inversely proportional to the usual sensitivity $\eta$ as defined for example in Ref.~\cite{barry2020sensitivity}. In the following we refer to $\Delta S$ as sensitivity, which we thus aim to maximize. The differential spectrum corresponding to Fig.~\ref{fig:Efield_opti}a is presented in Fig.~\ref{fig:Efield_opti}b, and the extrema of $\Delta S$ are presented in Fig.~\ref{fig:Efield_opti}c. Unlike in the weak $E$ case, we observe a non-monotonous sensitivity when increasing $B$, with even a drop to 0 expected around $30~\mathrm{G}$, and a maximal sensitivity obtained when the hyperfine transitions are merging. When the lines are merged, the sensitivity saturates. It is important to note that an increase by a factor 2 is also expected in the weak $E$ case when the hyperfine transitions are degenerate (e.g. for $B \gtrsim 50~\mathrm{G}$ with a 1 MHz linewidth), since the transition amplitude is also increased by a factor 2.

Finally, we note that for strong $E$ fields, i.e. case (iii), hyperfine transitions are degenerate for any value of $B$. In this case, Eq.~\ref{eq:transition_freq} is again relevant. A particular case is $d_\perp E \gg \Lambda$, for which Eq.~\ref{eq:transition_freq} can be approximated as $f_\pm(\vec{E}_\perp, \vec{B}_\perp) \approx D_{gs} + (2 \mp \cos{\phi}) \Lambda \pm d_\perp E_\perp$. In particular, the sensitivity is then completely independent of $\vec{B}$, both for its magnitude and direction.

We have seen that even with a single NV center, there exist regimes where analytical formulas can hardly be applied, and a numerical simulation approach might be necessary for sensitivity optimization. We now consider electric field sensing with NV ensembles.

Let us assume that a bias field $\vec{B}$ is applied orthogonal to a single NV orientation, e.g. $NV_1$. Thus, all other NVs have Zeeman splitting dominating over Stark shift, and their transition energies are well separated from the $E$-sensitive orientation, see Fig.~\ref{fig:Efield_opti}d. We consider a linewidth of 1 MHz, well-representative of NV ensemble. Fig.~\ref{fig:Efield_opti}(d-f) presents the transition amplitude and corresponding sensitivity calculations performed with same $\vec{B}, \vec{E}$ in $NV_1$ frame as in panels (a-c). To illustrate how the MW polarization degree of freedom can be exploited for sensitivity optimization, we now consider linearly polarized MW. The MW field is set orthogonal to $NV_1$ axis, with $\phi_{MW}=\pi/2$. The main difference with the single NV case is due to the contribution from VN-aligned centers, which results in up to 8 non-degenerate transitions just for $NV_1$ orientation, thus complicating the identification of each transition. Additionally, the polarized microwave causes an imbalance in amplitude between upper and lower frequency transitions, particularly visible for $B > 30~\mathrm{G}$. Eventually, the sensitivity to electric field, shown in Fig.~\ref{fig:Efield_opti}e, is completely different from the single NV case (Fig.~\ref{fig:Efield_opti}b). This illustrates that a thorough analysis with our code is needed in order to optimize the sensing scheme with NV ensembles. In this precise case, optimal sensitivity is reached for a magnetic field around $B=60~\mathrm{G}$. 
Around this value of $B$, the chosen MW polarization direction allows to selectively excite only the lower frequency transition, with transition strength twice as big as in the unpolarized case. The sensitivity, in turn, is increased by a factor 2. 
Interestingly, the microwave frequency for which maximum signal variation is expected is around 2880 MHz, between two hyperfine transitions.

\begin{figure}
	\centering
	\includegraphics[width=1\columnwidth]{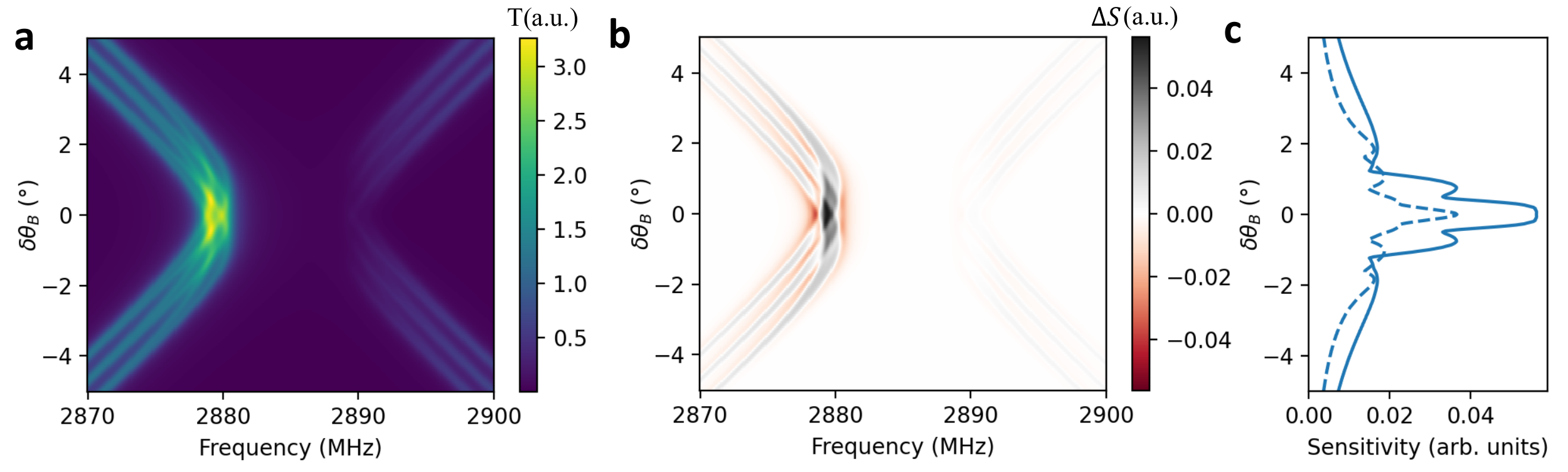}
	\caption{(a) Transition strength spectrum versus misalignement angle $\delta\theta_B$ away from the plane normal to $NV_1$ ($\theta_B = \pi/2$ in $NV_1$ frame). (b) Corresponding sensitivity $\Delta S$ versus $\delta\theta_B$. (c) Absolute maximum (solid) and minimum (dashed line) of $\Delta S$ versus $\delta\theta_B$. The configuration is identical to Fig.~\ref{fig:Efield_opti}(d-f), with $B=60~\mathrm{G}$}.
	\label{fig:Bmisalign}
\end{figure}

%As illustarted in Fig. \ref{fig:scanB}, the influence of misalignment is evident in small electric field regime as the applied magnetic field rotated.

To conclude this section, we briefly discuss the influence of bias magnetic field misalignment on the electric field sensing sensitivity. Fig.~\ref{fig:Bmisalign} illustrates how sensitivity is affected when $\vec{B}$ is rotated out of the plane normal to the NV axis. It is clearly seen that sensitivity drops quickly even for a few degree of misalignment. This is due to the Zeeman shift induced by the longitudinal component $B_z$ when $B$ is misaligned slightly out-of-plane (see Fig.~\ref{fig:Bmisalign}a). Note that this also illustrates that in an NV ensemble, the NV orientations different from the chosen $E$-sensitive one (i.e those with non-orthogonal bias magnetic field $\vec{B}$) are almost completely insensitive to electric field.

\section{Full vector electometry with a single ODMR spectrum}\label{sec:2nvorientations}

We now propose a new scheme for the full determination of the electric field vector using a single value of bias magnetic field $\vec{B}$.
The scheme considered in the previous section only allows to extract the magnitude of $\vec{E}$ in the plane normal to a single NV orientation. In theory, an ensemble ODMR spectrum, with up to 8x3 resonances, contains enough information to extract $\vec{E}$~\cite{broadway2018spatial}, but as already mentioned, the ratio $d_\parallel / d_\perp \ll 1$ leads to significant uncertainty in the longitudinal component $E_z$. Thus, the most common solution for vector electrometry is to use several directions of magnetic field bias $\vec{B}$: rotating $\vec{B}$ in the normal plane allows to identify the transverse vector direction, i.e polar angle $\phi_E$~\cite{dolde2011electric, doherty2014measuring, barson2021nanoscale}, and the longitudinal component $E_z$ can be obtained by aligning $\vec{B}$ orthogonal to a second NV orientation~\cite{yang2020vector}.
However, it can be cumbersome and time consuming to modify the bias magnetic field in experiments.
In Ref.~\cite{mittiga2018imaging}, the authors used another approach: they took advantage of the dependence of transition strength on linear microwave polarization to determine $\phi_E$, using the relation from Eq.~\ref{eq:imbalance}.

\begin{figure}
	\centering
	\includegraphics[width=0.7\columnwidth]{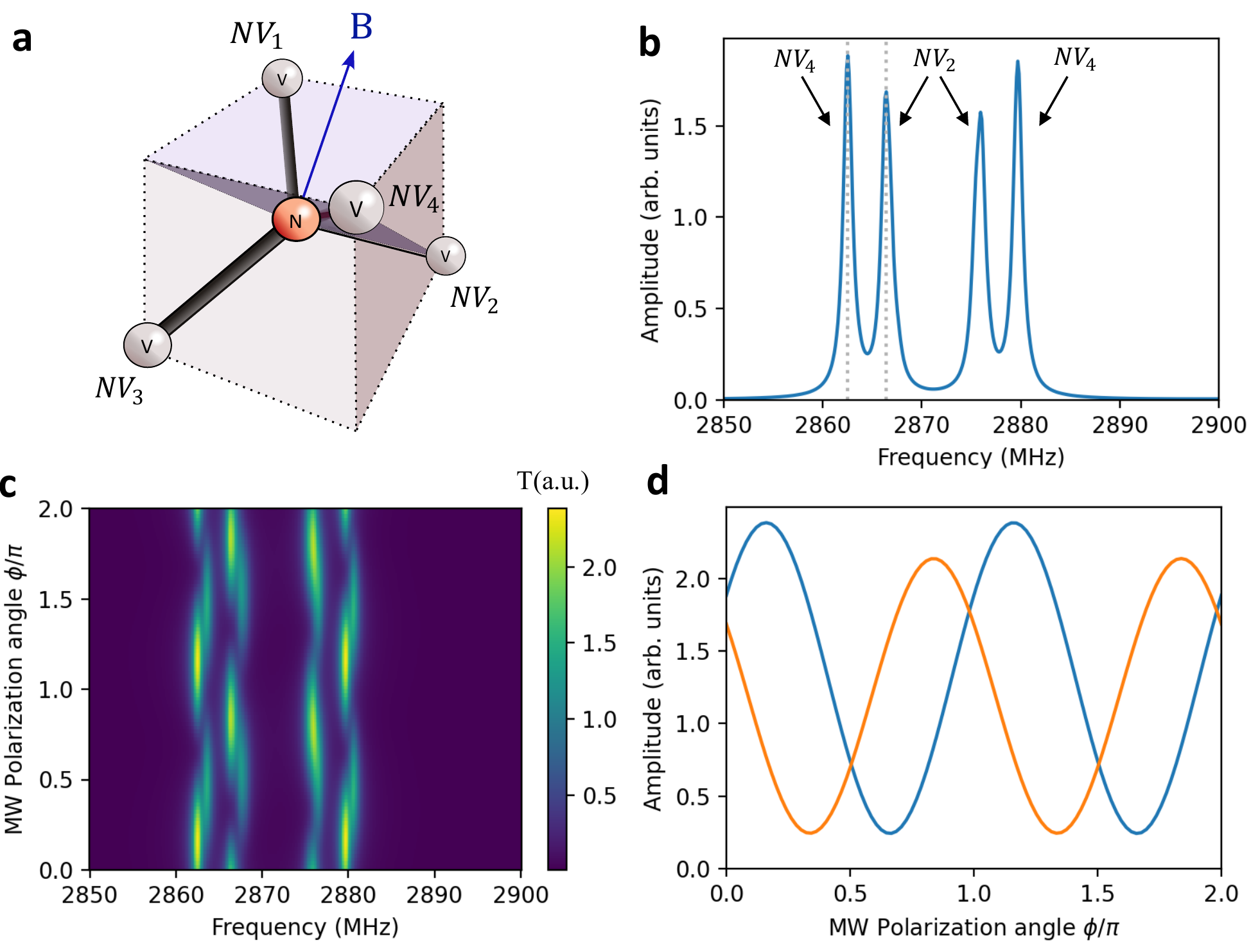}
	\caption{(a) Geometric configuration for our vector electrometry scheme. To illustrate our method, we choose a bias magnetic field $\vec{B}$ along the $<011>$ direction, normal to the plane containing $NV_4$ and $NV_2$. We set $B = 20~\mathrm{G}$, electric field $\vec{E} = (45, 15, -15) \cdot 10^6~\mathrm{V/m}$, in laboratory coordinate frame, and MW field orthogonal to $Z_L$ direction ($\theta_{MW} = \pi / 2$).  (b) Transition strength spectrum for linearly polarized MW with $\phi_{MW} =0$, where $\phi_{MW}$ is defined with respect to $X_L$. (c) Transition spectrum versus MW polarization angle $\phi_{MW}$. (d) Transition strength versus $\phi_{MW}$, at the lower frequency transition associated with $NV_4$ (blue) and $NV_2$ (orange), as identified in (b) by the two vertical dotted lines.}
	\label{fig:Bperp_2NVs}
\end{figure}

Our proposed scheme builds on their approach: the bias magnetic field $\vec{B}$ is chosen orthogonal to 2 NV orientations simultaneously, e.g., along $<011>$ as illustrated in Fig.~\ref{fig:Bperp_2NVs}a. Consequently, 2 NV orientations become $E$-sensitive ($NV_2$, $NV_4$). The ODMR spectrum presents two sets of resonance around 2870 MHz, due to the two different transverse components of $\vec{E}$ for the 2 chosen NV orientations, see Fig.~\ref{fig:Bperp_2NVs}b. The transition amplitude versus linear polarization angle $\phi_{MW}$ is presented in Fig.~\ref{fig:Bperp_2NVs}c, where the MW field is polarized orthogonal to $<001>$ direction. $\phi_{MW}=0$ corresponds to polarization along $X_L$ in the lab frame, $\phi_{MW}=\pi/2$ along $Y_L$, and so on. For two transition energies associated with the two different NV orientations, the evolution of transition amplitude is periodic, with an offset that is different in the two cases. Note that the rotation of the MW polarization is not normal to any of the 2 NV axis, so strictly speaking the dependence on $\phi_{MW}$ is not sinusoidal. Nevertheless, the simulation allows to extract $\phi_E$ for the two transverse projections. 
For example, in Fig.~\ref{fig:Bperp_2NVs}d, the maximal transition amplitude for $NV_4$ (blue line) is at $\phi_{MW} = \pi/6$. Using Eq.~\ref{eq:imbalance}, and taking into account the correct offset in $\phi_{MW}$ due to our arbitrary choice of $\phi_{MW}=0$ along $X_L$, we get $\phi_E = \pi / 3$, which is precisely the value of $\phi_E$ in the frame of $NV_4$ for the chosen $\vec{E} = (45, 15, -15) \cdot 10^6~\mathrm{V/m}$ (expressed in laboratory coordinate frame). Having determined the amplitude and direction of $\vec{E}_\perp$ in one NV frame, the only unknown is $E_z$ in the same NV frame: we have restricted the possible solutions for vector $\vec{E}$ to a single line. 
Eventually, the knowledge of the transverse projection vector in the two NV axes combines into a unique vector $\vec{E}$, at the intersection of the two lines of solution associated with the two axes. 
Thus, the electric field vector can be determined unambiguously. The advantage of our methods is that, with proper design of the MW antenna, rotation of the linear MW polarization can be achieved all-electrically~\cite{staacke2020method}, with much faster timescale than a change of magnetic field.

\section{Conclusion}
To conclude, in this work we developed an open-source Python based code (accessible at \url{https://github.com/chris-galland/NV-ODMR-simulation}) to compute the NV electronic energy levels and the corresponding ODMR spectrum in presence of external magnetic and electric fields. %, from both a single NV center and an NV ensemble. 
This code is a convenient tool to explore how the ODMR spectrum varies with different parameters of interest (e.g. fields intensity and direction or linewidth), for both single NV and ensembles of them. %The sweeping variables and range can be easily adapted to the user requirements. 
In particular, having the NV sensing application in mind, we focused on the sensitivity computation. Optimizing the sensitivity with respect to the experimental conditions is not always straightforward: even with a single NV center, there exist regimes where analytical formulas
can hardly be applied, and a numerical simulation is necessary. With our code we try to meet this need. In particular, sensitivity maps can be simulated under varying experimental parameters, thus facilitating the working point optimization and allowing for the development of new sensing schemes.
%Note that, given the breadth of the subject and its complexity, it's not possible to include all possible elements into the simulation, however, the code we provide offer a solid basis for future extensions, tailored on the user requirement. For example, the effect of $^{15}N$ or the impact of strain in the diamond lattice could be included.

In the second part of the paper, we showed how our code can provide new insights into electric field sensing and we suggested a novel electrometry scheme, which relaxes the effort on dynamic bias magnetic fields alignment. 
On the one hand, we discussed electric field sensing in different regimes, presenting sensitivity maps and exploring the impact of experimental imperfections, like the misalignement of the bias magnetic field. %electric field sensing is quite complex, this approach can be useful for a newcomer to be successful 
On the other hand, we showed how full vector electrometry using an NV ensemble is possible without having to dynamically re-align the bias magnetic field in different directions. Our method is based on the change of ODMR amplitudes with MW polarization angle and only requires a properly designed antenna. This method will allow to significantly decrease the experimental complexity and speed up the measurement. Finally, we believe that our open source code will help students and researchers explore the physics of NV center ensembles, optimize quantum sensors based on them, and generalise the simulation to other color centers.

\subsection*{Acknowledgement}
This project has received funding from the Swiss National Science Foundation (grants No. 198898 and 204036) and from EPFL Interdisciplinary Seed Fund.

\bibliography{sample}

\end{document}